\documentclass[twocolumn,
prd,aps,superscriptaddress,tightenlines]{revtex4}
\usepackage{pstricks}

\newif\ifpdf
\ifx\pdfoutput\undefined
\pdffalse 
\else
\pdfoutput=1 
\pdftrue
\fi

\def\OMIT#1{{}}
\def\lqcd{\Lambda_{\rm QCD}}

\newcommand{\nn}{\nonumber}
\newcommand{\beq}{\begin{equation}}
\newcommand{\eeq}{\end{equation}}
\newcommand{\beqa}{\begin{eqnarray}}
\newcommand{\eeqa}{\end{eqnarray}}

\newcommand{\bn}{{\bar n}}
\newcommand{\bnP}{\bar {\cal P}}
\newcommand{\nP}{{\cal P}}
\def\nslash{n\!\!\!\slash}
\def\bnslash{\bar n\!\!\!\slash}

\def\abs#1{ \left| #1 \right| }

\def\diff{{\text d}}

\begin{document}
\ifpdf
\else
\fi

\title{Enhanced nonperturbative effects in jet distributions}

\author{Christian W.~Bauer}
\affiliation{Department of Physics, University of California at San Diego,
  La Jolla, CA 92093\vspace{4pt} }

\author{Aneesh V.~Manohar}
\affiliation{Department of Physics, University of California at San Diego,
  La Jolla, CA 92093\vspace{4pt} }

\author{Mark B.~Wise}
\affiliation{California Institute of Technology, Pasadena, CA 91125\vspace{4pt} }

\begin{abstract}
We consider the triple differential distribution $\diff \Gamma/\diff E_J \diff m_J^2 \diff \Omega_J$ for two-jet events at center of mass energy $M$, smeared over the endpoint region  $m_J^2 \ll M^2$, $\abs{2 E_J -M} \sim \Delta$, $ \lqcd \ll \Delta \ll M$.
The leading nonperturbative correction, suppressed by $\lqcd/\Delta$, is given by the matrix element of a single operator. A similar analysis is performed for  three jet events, and the generalization to any number of jets is discussed. At order $\lqcd/\Delta$, non-perturbative effects in four or more jet events are completely determined in terms of two matrix elements which can be measured in two and three jet events.
\end{abstract}

\maketitle

This paper studies the semi-inclusive two-jet distribution $\diff \Gamma/(\diff E_J \diff m_J^2 \diff \Omega_J)$ in high energy processes such as $Z \to q \bar q$, where $E_J$ and $m_J$ are the energy and invariant mass of one of the jets, $J$. No restriction is placed on the kinematics of the second jet, $J^\prime$. We are interested in non-perturbative effects that are enhanced in the endpoint region $\abs{2E_J - M} \ll M$,  $m_J^2 \ll M^2$. In the endpoint region the jet $J$ is narrow, and has energy close to $M/2$. At leading order in the parton model description, the quark and antiquark each have energy $M/2$, and $m_J^2 = 0$. Gluon radiation and non-perturbative effects such as hadronization can cause $E_J$  and $m_J^2$ to deviate from their leading order parton model values. The aim of this letter is to characterize these non-perturbative effects in QCD in terms of operator matrix elements and study the relationship between nonperturbative effects in two jets, and events with three or more jets. These non-perturbative effects do not depend on the precise jet definition; for example, a suitable choice is the Sterman-Weinberg prescription~\cite{SW}. 

In the endpoint region, the decay distribution for two jet events can be written as
\beqa\label{doublefactor1}
&&\frac{\diff \Gamma}{\diff E_J \diff m_J^2 \diff \Omega_J} = T(M,\Omega_J) J(M,m_J^2/M) \\
&&\times\int \diff k^+ J(M, k^+) S\left(2E_J-M-k^+-m_J^2/\left(2E_J\right)\right).\nn
\eeqa
Here the hard function $T$ and the jet function  $J$ are calculable perturbatively and $S$ is a nonperturbative shape function. The precise form of the perturbative expansions of these functions depends on the jet definition used. In the endpoint region this distribution must be smeared over a suitable region of $E_J$ and $m_J^2$ to be physically meaningful.  This general  form for the decay distribution has  been obtained ~\cite{EventShapes} previously using standard ``diagrammatic'' factorization methods~\cite{review}. We begin this paper by reviewing the derivation of Eq.~(\ref{doublefactor1}) using soft collinear effective field theory (SCET).

The effective field theory SCET \cite{scet1,bs,bpssoft} is appropriate for the kinematic region of interest. SCET describes the interaction of collinear and ultrasoft (usoft) degrees of freedom with momenta scaling as $p_c = (n \cdot p, \bn \cdot p, p^\perp) \sim M(\lambda^2, 1, \lambda)$ and $p_{us} \sim M(\lambda^2, \lambda^2, \lambda^2)$.  The light-like vectors $n$ and $\bn$ satisfy $n^0=\bn^0=1$ and $\mathbf{n}=-\bar\mathbf{n}$ and the perpendicular components of any four vector $V$ are defined by  $V^{\mu}_{\perp}=V^{\mu}-(n \cdot V)\bar n^{\mu}/2-(\bar n \cdot V)n^{\mu}/2$. For our analysis, $\lambda^2 \sim \Delta /M \ll  1$. 
The effective theory provides a simple method for the factorization of hard, collinear and usoft degrees of freedom at the operator level. For example, the factorization of usoft degrees of freedom arises because they can be decoupled from the collinear degrees of freedom using a simple field redefinition. SCET gives field theoretical definitions of the various ingredients in Eq.~(\ref{doublefactor1}).

The first step in the SCET derivation of  Eq.~(\ref{doublefactor1}) is  matching the full theory current $j^\mu$ onto SCET.  The current in the effective theory at leading order in $\lambda$ is
\beqa
{j}^\mu = [\bar \xi_\bn W_\bn] \Gamma^\mu C(\nP^\dagger, \bnP) [W^\dagger_n \xi_n]\,,
\label{jdef}
\eeqa
where $\Gamma^\mu = g_V \gamma^\mu_\perp + g_A \gamma^\mu_\perp \gamma_5$, $g_{V,A}$ are the vector and axial couplings of the quarks to the $Z$ boson, and $C(\nP^\dagger, \bnP) $ is the matching coefficient which is one at tree level. The field $\xi_n$ denotes a collinear fermion in the $n$ direction, and we have used the convention
\begin{eqnarray}
\xi_n(x) = \sum_{\tilde p} e^{-i \tilde p \cdot x} \xi_{n, \tilde p}(x),
\label{bsdef}
\end{eqnarray}
where $\tilde p$ is the label momentum which contains components of order $1$ and order $\lambda$. The order $\lambda^2$ components are associated with the spacetime dependence of the fields.

The label operators $\nP$, $\bnP$ pick out the order one momenta of the collinear fields and $W_n(x)$ denotes a Wilson line of collinear gluons along the path in the $\bn$ direction. The Wilson lines $W_{n, \bn}$ are required to ensure gauge invariance of the current in the effective theory \cite{bs}. The Lagrangian of the effective theory does not contain any direct coupling of collinear particles moving in the two separate directions defined by $n$ and $\bn$ \cite{bfprs}, however they can still interact with one another via the emission of usoft gluons. The coupling of collinear and usoft degrees of freedom in the Lagrangian can be eliminated via the BPS field redefinition \cite{bpssoft}
\beqa
\xi_n \to Y_n^{\dagger} \xi_n \,, \qquad A_n \to Y_n^\dagger A_n Y_n
\eeqa
where 
\beqa
Y_n(z) = \exp \left[ i g \int \diff s \,n \cdot A_{us}(n s + z) \right]
\label{yline}
\eeqa
denotes a path-ordered Wilson line of usoft gluons in the $n$ direction from $s=0$ to  $s=\infty$,  since we are dealing with final state collinear fields. [For annihilation, $Y_n$ is from $s=-\infty$ to $s=0$.] It is  well known from the ``diagrammatic''  approach to factorization that the usoft degrees of freedom couple to collinear degrees of freedom via a Wilson line \cite{pink}.

To derive the factorized form in Eq.~(\ref{doublefactor1}), we start from the general expression
\beqa\label{fulldiff}
{\diff  \Gamma \over \diff^4 r} &=& {1 \over 2M} \sum_{JX} (2 \pi)^4 \delta^4(M-p_J-p_X) 
\delta^4(r-p_J)\nn \\
&&\qquad \times \left|\epsilon_\mu\left\langle J,X | j^\mu | 0 \right\rangle \right|^2 \,,
\eeqa
where the sum includes the phase space integrations over all the particles in the final state, $j^\mu$ denotes the current producing the $q \bar q$ pair, and $\epsilon_\mu$ is the polarization of the decaying particle. The final state hadrons have been divided into those in the quark jet $J$, with total momentum $r$ and the remaining hadrons (including the antiquark jet) which form $X$.

Using these definitions, the matrix element of the current in Eq.~(\ref{fulldiff}) becomes
\beqa
\left\langle JX | j^\mu | 0 \right\rangle &=& \sum_{\tilde q_1 \tilde q_2} C(q_1^+,  q_2^-) \left\langle J |
T \Bigl\{[\bar \xi_\bn W_\bn]^{a}_{\tilde q_1 \alpha }\Bigr\} | 0 \right\rangle
\left( \Gamma^\mu\right)_{\alpha\beta}
\nn\\[-5pt]
&&
\times
\left\langle X \Bigl| T
\Bigl\{ [Y_\bn Y_n^{\dagger} ]_a{}^b [W^\dagger_n \xi_n]_{\tilde q_2 \beta b }\Bigr\} \Bigr| 0 \right\rangle\,,  
\label{eq9}
\eeqa
where $\alpha,\beta$ $(a,b)$ denote spin (color) indices, and the subscript $[\cdot ]_{\tilde q}$ denotes the total label momentum of the operator inside. In Eq.~(\ref{eq9}), and in the subsequent equations, we will use the notation $n \cdot p \equiv p^+$, $\bar n \cdot p \equiv p^-$ for any 
four-momentum $p$.

The operators inside the matrix elements in Eq.~(\ref{eq9}) are time-ordered. The operator $T\bigl\{ Y_n \bigr\}$ is a Wilson line where the operator  time-ordering agrees with the path-ordering of matrix multiplication, $T\bigl\{ Y_n \bigr\}=Y_n$. The operator $T \bigl\{ Y_n^\dagger \bigr\}$ has operator time-ordering in the opposite order as the  path-ordering of matrix multiplication. The two orderings can be made to coincide by taking the transpose of all the matrix indices, so that  $T \bigl\{ Y_{n a}^{\dagger }{}^{b} \bigr\}$ is the Wilson line in the $\mathbf{\bar 3}$ representation, i.e., $T \bigl\{ Y_{n a}^{\dagger}{}^{b} \bigr\}= \overline Y_{n}{}^{b}{}_a $. Similar results hold for the anti-time ordered products.

Using Eq.~(\ref{eq9}), the differential distribution becomes
\beqa\label{fulldiff1}
{\diff \Gamma \over \diff^4 r} &=& {1 \over 2 M}\!\int  \!\! \diff^4\!s\ (2 \pi)^4 \, \delta^4(M-r-s) 
\nn\\
&&\quad
\epsilon_\mu \epsilon^{\nu *}\left( \Gamma^\mu\right)_{\alpha\beta}\left(\bar \Gamma_\nu\right)_{\rho\sigma}
\left| C(r^+,s^-) \right|^2
\nn\\
&&\!\!\!\!
\times \left[ \sum_X \delta^4(s-p_X) \left|\left\langle X | T\Bigl\{[ Y_\bn Y_n^\dagger] [W_n^\dagger \xi_n]_{\tilde s} | 0 \right\rangle \Bigr\} \right|^2\right]_{\beta\rho a}^{\quad\ c}
\nn\\
&&\!\!\!\!
\times \left[ \sum_J \delta^4(r-p_J) \left|\left\langle J |T \Bigl\{ [\bar \xi_\bn W_\bn]_{\tilde r} | 0 \right\rangle
\Bigr\} \right|^2 \right]_{\sigma\alpha c}^{\quad\ a}.
\label{9a}
\eeqa
The two terms in square bracket can be simplified, following Ref.~\cite{bpssoft}. 
For the second term we have
\beqa
\sum_J &&\!\!\!\!\!\!\!\!\!\!\! \int  \diff^4z \, e^{i (r-p_J)\cdot z}\left[ \left| \left\langle J |
T \Bigl\{ [\bar \xi_\bn W_\bn]_{\tilde r}  | 0 \right\rangle \Bigr\} \right|^2 \right]_{\sigma\alpha c}^{\quad\ a}
\nn\\
&=& 
\left( \frac{\bnslash}{2} \right)_{\sigma \alpha}\delta_c^a\ J_{\tilde r}(r^-) ,
\label{jet1}
\eeqa
where the last equality defines $J_{\tilde r}(r^-)$.

\begin{figure}
\caption{Pictorial representation of the Wilson lines occurring in the shape function $S(k)$ defined in Eq.~(\ref{sdef}). \label{fig}}
\psset{unit=0.3cm}
\begin{pspicture}(0,-1)(7,6)
\psline(0,0)(2.3,2.3)
\psline[arrowscale=2]{<-}(2.3,2.3)(5,5)
\psline[arrowscale=2]{->}(0,0)(1.2,-1.2)
\psline(0.8,-0.8)(2,-2)
\psline(7,3)(2,-2)
\psline[arrowscale=2]{<-}(4.7,0.7)(2,-2)
\rput(-0.5,-0.5){$u$}
\rput(1.5,-2.5){$0$}
\rput(2,3){$\bn$}
\rput(0.5,-1.5){$n$}
\rput(5,0){$\bn$}
\end{pspicture}
\end{figure}
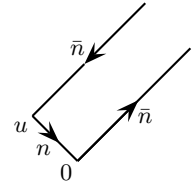
Combining the first term with the $\delta_c^a$ in Eq.~(\ref{jet1}) gives
\beqa
\sum_X &&\!\!\!\!\!\!\!\!\!\!\! \int \! \diff^4z \, e^{i (s-p_X)\cdot z} \left[ 
\left| \left\langle X | T
\Bigl\{ [ Y_\bn  Y_n^\dagger] [W_n^\dagger \xi_n]_{\tilde s} \Bigr\} | 0 \right\rangle \right|^2\right]_{\beta\rho a}^{\quad\ a}
\nn\\
&=&
3  \int \diff k^+ J_{\tilde s}(k^+) S(-s^+-k^+)  \left( \frac{\nslash}{2} \right)_{\beta\rho}\,.
\eeqa
The shape function $S(k^+)$ is defined by~\cite{EventShapes}
\beqa
S(k) &=& {1\over 3} \int \!{\diff u\over 2 \pi}\ e^{i k u}\left \langle 0 \left| \text{Tr}\ \overline T \Bigl\{ [Y_n Y_\bn^\dagger](nu)\Bigr\} T \Bigl\{ [Y_\bn Y_n^\dagger ](0) \Bigr\} \right| 0 \right\rangle 
\nn\\
&=&{1\over 3} \left \langle 0 \left|{\overline  Y}_{n }^{\dagger c}{}_a   Y_{\bn c}^{\dagger }{}^ b
 \delta(k - i n \cdot \partial) Y_\bn{}_b{}^ e    {\overline  Y}_n{}^a{}_e    \right| 0 \right\rangle .
\label{sdef}
\eeqa
The Wilson loop, shown in Fig.~\ref{fig}, is regulated to preserve reparametrization invariance~\cite{rpi}, and is a function of $\bn \cdot n u$.
$S(k)$ is normalized so that
$\int_{-\infty}^\infty \diff k\ S(k) = 1.$

The jet function in Eq.~(\ref{jet1}) depends on $\tilde r$, which has both $+$ and $\perp$ components. One can align $\bn$  with the jet-axis, i.e.\ choose ${\mathbf r}_\perp=\mathbf{0}$. With this choice, we write
\beqa
J(r^+,r^-) = J_{r^+,\mathbf{0}_\perp}(r^-).
\label{jet2}
\eeqa
Combining Eqs.~(\ref{9a})--(\ref{jet2}) and using $\diff ^4 r = (M/4) \diff E_J \diff m^2_J \diff \Omega_J$ gives Eq.~(\ref{doublefactor1}) with the hard function defined by
\beqa
T(M,\Omega_J) = {3\over 512 \pi^4} |C(M,M)|^2 \epsilon_\mu \epsilon^{\nu *} \text{Tr} \left[ \Gamma^\mu \nslash \bar \Gamma_\nu \bnslash \right].
\eeqa

The simple form for the shape function in Eq.~(\ref{sdef}) arises because the decay distribution is totally inclusive on the unobserved usoft degrees of freedom and on the other jet $J^\prime$. This would not be the case for the thrust distribution near $T=1$, since the deviation of $T$ from unity depends on the momentum of the usoft degrees of freedom. 

We can simplify the discussion by restricting ourselves to the leading order in perturbation theory, where
\beqa
C(M,M) = 1, \quad J(M,k^+) =2 \pi \delta(k^+).
\label{lo}
\eeqa
Perturbative corrections to this result can be included by calculating the two jet functions $J$ and the Wilson coefficient $T$ to higher orders in perturbation theory.
For Sterman-Weinberg jets~\cite{SW}, the perturbative corrections contain logarithms of the cone angle $\delta$ and the minimum energy cut $E_c$. In the effective theory, logarithms of $\delta$ arise for precisely the same reason as in the full theory; one has to separate three collinear-field final states into two-jet and three-jet events. Logarithms of $E_c$ arise in the effective theory from the separation into soft and collinear fields. One can sum these logarithms using renormalization group methods in the effective theory.

Integrating Eq.~(\ref{doublefactor1}) over $m_J^2$ using Eq.~(\ref{lo}) gives
\beqa\label{difffinal2}
\frac{\diff \Gamma}{\diff E_J \diff \Omega_J }  =2 {\diff \Gamma^{(0)} \over \diff \Omega_J}
S(2E_J-M) + \ldots
\eeqa
where $\diff \Gamma^{(0)}/\diff \Omega_J$ is the parton model differential decay rate and the ellipsis denotes subdominant perturbative and power corrections.
To this order the shape of the jet energy spectrum is determined entirely by the non-perturbative shape function. Non-perturbative usoft radiation is not preferentially directed along the jet directions and  will reduce the energy associated with the jets. Thus, $S(k)=0$ for $k>0$.    

To proceed further we consider observables in which we smear the jet energy distribution over a region $|2E_J-M| \sim \Delta \ll M$, with $\Delta \gg \lqcd$~\footnote{Smearing Eq.~(\ref{doublefactor1}) over an energy region of width $\Delta$ misses corrections suppressed by powers of $\Delta/M$. For example,  operators with $\mathbf{D}_\perp$ are suppressed, and so are dropped in
the definition of the jet function in SCET. These effects are calculable perturbatively and  can easily be included.}. In this region there are enhanced non-perturbative corrections suppressed only by powers of $\lqcd/\Delta$. Note that for the total rate the non-perturbative corrections are much smaller since they are suppressed by  powers of $\lqcd/M$. To be more precise, consider a smooth window function $w_{\Delta}(E_J)$ which when integrated over $E_J$ is normalized to unity and has support concentrated in the region $|E_J-M/2|<\Delta$. Then,
\beq
\label{smeared}
\left(\frac{\diff \Gamma }{\diff \Omega_J }\right)_{\Delta} =\int \!\! dE_J \, w_{\Delta}(E_J)\frac{\diff \Gamma}{\diff E_J \diff \Omega_J }, 
\eeq
is such an observable. For it we can expand the shape function in a power series, given by
\beqa\label{Sexp}
S(k)  = \delta(k) 
- \delta'(k)  \, \langle 0 |   O_1 | 0 \rangle + \frac{1}{2} \delta''(k)   \, \langle 0 |  O_2
| 0  \rangle + \ldots \,
\eeqa
where
\beqa
O_m =  {1\over 3}  {\rm Tr} \left[ Y^\dagger_\bn
  (i n \cdot D)^m  Y_\bn\right] 
\label{18a}
\eeqa
We can rewrite $O_1$ as
\beqa
O_1 = O_1^q  = {1\over 3}
 {\rm Tr}\,  \int^{\infty}_0 \!\! \diff s \, {\cal G}_{n\bn}(\bn s)\ ,
\label{18}
\eeqa
where
\beqa
{\cal G}_{n_1  n_2}(\bn s) = Y_\bn(\bn s,0)^\dagger  n_1^\mu n_2^\nu G_{\mu \nu} Y_\bn(\bn s,0).
\eeqa
Here, $G^{\mu\nu}$ is the gluon field-strength (with a factor of the strong coupling absorbed into it) and $Y_\bn(\bn s,0$) denotes a usoft Wilson line  along the $\bn$ direction from $0$ to $\bar n s$. 
Note that the Fourier transform of $S(k)$ is a function only of the combination $\left(\bar n \cdot n \right)u$, so
the matrix element of $O_m$ must have the form
\beqa
\langle 0 | O_m | 0 \rangle = \left(\bar n \cdot n \right)^{m} \mathcal{A}_{m}^q, 
\label{omdef}
\eeqa
where $\mathcal{A}_{m}^q$ is a number of order $\lqcd^{m}$. If the observed jet is the antiquark jet, then $in \cdot \partial$ in Eq.~(\ref{18}) is replaced by $i \bn \cdot \partial$, and the leading correction is given by the vacuum matrix element of
the antiquark operator
\beqa
O_m^{\bar q} = {1\over 3} {\rm Tr}\, [\overline Y_n^{\dagger}  (i \bn \cdot D)^m \overline Y_n],
\eeqa
which is $\left( \bn \cdot n \right) \mathcal{A}_m^{\bar q}$. By charge conjugation invariance, $\mathcal{A}_m^{\bar q}= \mathcal{A}_m^{q}$.

The nonperturbative corrections to the differential decay rate in Eq.~(\ref{difffinal2}) are singular at $E_J=M/2$. This is also the case for the perturbative corrections. However for the smeared observable in Eq.~(\ref{smeared}) the expansion of the shape function gives,
\beqa
\label{smeared1}
\left(\frac{\diff \Gamma }{\diff \Omega_J }\right)_{\Delta}
\!\!\!=
\left(\frac{\diff \Gamma^{(0)} }{\diff \Omega_J }\right) \bigg[1+ w^{\prime}_{\Delta}(M/2) \mathcal{A}_1^q 
+\ldots \bigg].
\eeqa    
 Since the matrix elements of the operators scale as $
\mathcal{A}_m^q \sim \lqcd^m$ and the $m$'th derivative of the window function (evaluated at $E_J=M/2$) scales as $1/\Delta ^m$, the square brackets on the right hand side of Eq.~(\ref{smeared1}) contain an expansion in powers of  $\lqcd/\Delta$.

A similar calculation also applies for events with more than two jets in the final state. 
For example, three jet events contain an energetic gluon radiated off one of the quarks at large angle. The subscripts $1,2,3$ refer to the quark, gluon and antiquark jets, respectively. One of the jets is unobserved, and is summed over. Consider the case where the antiquark jet (jet 3) is unobserved, and the quark and gluon jet are observed.
The light like vectors $n_{1,2}$ are defined by $n_1=(1,\mathbf{n}_1)$, $n_2=(1,\mathbf{n_2})$ where $\mathbf{n}_{1,2}$ are unit vectors in the direction of the two observed jets. One can the construct the third vector
\beqa
n_3^0 = 1\,,\qquad \mathbf{n}_3 = - \frac{E_1 \mathbf{ n}_1 + E_2 \mathbf{ n}_2}{\left | E_1\mathbf{ n}_1 + E_2 \mathbf{ n}_2 \right|}\, ,
\eeqa
using only information from the two observed jets. A similar argument holds if the quark jet, or the gluon jet, is the unobserved jet. 

Let $i$ and $j$ denote the two observed jets, and $r$ denote the unobserved jet, where $(i,j,r)$ is some permutation of $(1,2,3)$.
We find at tree level that
\beqa
\frac{\diff \Gamma_{ij}}{\diff E_i \diff\Omega_i  \diff E_j  \diff \Omega_j} 
&=& \frac{ \diff \Gamma_{ij}^{(0)}}{ \diff E_i d\Omega_i  \diff \Omega_j} n_r \cdot n_j \label{threejet}\\[5pt]
&&\times S_{ij}\left(n_r\cdot n_i E_i +  n_r\cdot n_j E_j -M \right),\nn
\eeqa
for the differential decay distribution for the observed jets $i$ and $j$ in terms of a non-perturbative shape function $S_{ij}$.

In the three-jet case, the current in SCET contains  the gluon field $A_{n_2}$. Factoring the usoft degrees of freedom results in the  shape function 
\begin{widetext}
\beqa\label{s12def}
S_{ij}(k)  &=&{1\over 4} \int \! {\diff u\over 2 \pi}\ e^{i k u }{\rm Tr} \left \langle 0
\left| \overline  T \Bigl\{  [Y_{n_3} Y_{n_2}^\dagger  T^A Y_{n_2} Y_{n_1}^\dagger ](u\, n_r) \Bigr\} T \Bigl\{
[Y_{n_1} Y_{n_2}^\dagger T^A Y_{n_2} Y_{n_3}^\dagger](0)
\Bigr\}  \right| 0 \right\rangle \ .
\eeqa 
\end{widetext}
Thus, a different shape function determines the usoft physics in three-jet events, and in general a new nonperturbative function is required for each additional jet. This is not surprising since the color structure is very different for events with different numbers of jets.  The time-ordered product $T \bigl\{ Y_{n a}^{\dagger}{}^b Y_{n c}{}^d \bigr\}$
is equal to $(1/3)\delta_a^d \delta_c^b+2\mathcal{Y}_{n A B}\left( T^A \right)_c{}^b \left(T^B \right)_a{}^ d$ where $\mathcal{Y}_n$ is the adjoint Wilson line from $0$ to $\infty$ in the $n$ direction.

In the kinematic region $|M-E_i n_r\cdot n_i -E_j n_r\cdot n_j| \sim \Delta$ an expansion of $S_{ij}(k)$ analogous to the one in Eq.~(\ref{Sexp}) can be performed, and the first correction to the smeared rate is determined by the operator
\beqa
O_1^{(3)} &=&  \frac{1}{3}{\rm Tr}\ Y_{n_1}^\dagger \left(i n_r \cdot D \right) Y_{n_1}  +  \frac{1}{3}{\rm Tr}\ \overline Y_{n_3}^\dagger \left(i n_r \cdot D \right) \overline Y_{n_3}\nn\\
&&+ \frac{1}{8}{\rm Tr}\ \mathcal{Y}_{n_2}^\dagger\left(i n_r \cdot D \right) \mathcal{Y}_{n_2}. 
\eeqa
The matrix element of this operator is given by
\beqa
\langle 0 | O_1^{(3)} | 0 \rangle&=&  n_r \cdot (n_1+ n_3)\, \mathcal{A}_1^q 
+ n_r \cdot n_2 \, \mathcal{A}_1^g .\,\,
 \label{3jetqmatrix}
\eeqa
where $\mathcal{A}_1^q$ is the same number that occurs in two-jet events, and is defined by Eq.~(\ref{omdef}) and $A_1^g$ is a new number of order $\lqcd$.

In three jet events with an unobserved gluon jet, $n_r=n_2$ and the vacuum matrix element of $O_1^{(3)}$ is $n_2\cdot\left( n_1 + n_3 \right) A_1^q$, and is completely determined by the two jet case. If the unobserved jet is the quark jet, then $n_r=n_1$, and the vacuum matrix element of $O_1^{(3)}$ is $ \left(n_1 \cdot n_2\right)A_1^g+ \left(n_1 \cdot n_3\right)A_1^q$, and if the unobserved jet is the antiquark jet, the vacuum matrix element of $O_1^{(3)}$ is $ \left(n_3 \cdot n_2\right)A_1^g+ \left(n_1 \cdot n_3\right)A_1^q$. If one cannot distinguish quark and gluon jets, then the three jet distributions are given by averaging over the unobserved jet being the quark, gluon or antiquark jet.

 One can repeat the above analysis for events with four or more jets. In the endpoint regions, the generalized shape functions can be expanded as in Eq.~(\ref{Sexp}), and the leading non-perturbative corrections are expressed in terms of $\mathcal{A}_1^q$ and $\mathcal{A}_1^g$ which characterize two and three-jet events. An important feature is that, even though the shape functions for events with different numbers of jets are very different, at order $\Delta/\lqcd$ no new non-perturbative quantities enter for four or more jets \emph{after smearing} by an amount $M\gg \Delta \gg \lqcd$ over the endpoint region. The reason is that the only Wilson lines that can arise from hard quark  anti-quark and gluon radiation are in the
$\mathbf{3}$, $\mathbf{\bar 3}$ and $\mathbf{8}$ representations, and these already occur in the two and three jet cases. One can also predict the distributions for two jets plus a hard photon in terms of $\mathcal{A}_1^q$ measured in two jet events, since the adjoint Wilson line from gluon emission does not enter.

$\mathcal{A}_1^q$ and $\mathcal{A}_1^g$ can be determined by fitting to the experimental data on two and three-jet events. They can also be computed numerically by light-cone lattice gauge theory methods~\cite{lattice}, since they only involve Wilson lines along light-like directions. 

This work was supported in part by the Department of Energy under grants DOE-FG03-97ER40546 and DOE-FG03-92ER40701. We would like to thank S.~Ellis, G.~Korchemsky, G.~Sterman, and I.W.~Stewart for helpful discussions.


\end{document}